\documentstyle[preprint,aps]{revtex}
\begin{document}
\draft
\title{Radiation Linewidth of a Long Josephson Junction \\ 
in the Flux-Flow Regime}
\author{A. A. Golubov \thanks{
Permanent address: Institute of Solid State Physics, Russian Academy of
Sciences, Chernogolovka, Moscow district, 142432, Russia }, 
B. A. Malomed \thanks{
Permanent address: Department of Applied Mathematics, School of Mathematical
Sciences, Tel Aviv University, Ramat Aviv 69978, Israel }, 
and A. V. Ustinov}
\address{Institute of Thin Film and Ion Technology\\ Research Center
(KFA), D-52425 \\J\"ulich, Germany}
\maketitle

\begin{abstract}
Theoretical model for the radiation linewidth in a multi-fluxon state of a
long Josephson junction is presented. Starting from the perturbed
sine-Gordon model with the temperature dependent noise term, we develop a
collective coordinate approach which allows to calculate the finite
radiation linewidth due to excitation of the internal degrees of freedom in
the moving fluxon chain. At low fluxon density, the radiation linewidth is
expected to be substantially larger than that of a lumped Josephson
oscillator. With increasing the fluxon density, a crossover to a much
smaller linewidth corresponding to the lumped oscillator limit is predicted.
\end{abstract}

\newpage
\narrowtext

A long Josephson junction is an example of a distributed nonlinear
oscillator which time-dependent response is associated with its intrinsic
spatial dynamics. An externally applied magnetic field $H$ penetrates into
the junction in form of Josephson fluxons (solitons), each of them carrying
one magnetic flux quantum $\Phi_0$. As schematically shown in Fig.~\ref
{FFS:schematics}, fluxons move across the junction under the influence of
the bias current generating an electromagnetic radiation at the junction
boundary. The frequency $f$ of the radiation is given by the Josephson
relation $f=V/\Phi_0$, where $V$ is a dc voltage induced by the fluxon
motion.

For a lumped (short) Josephson junction, the linewidth $\delta f$ of the
emitted radiation is determined by thermal fluctuations of current passing
through the junction. Assuming a Nyquist noise spectrum, for the
current-biased lumped Josephson tunnel junction the full linewidth at half
power is given by expression \cite{Stephen:68,Dahm:69}

\begin{equation}
\label{JJ-linewidth}\delta f\equiv \Delta f_T\cdot f=\frac{4\pi k_{{\rm B}}T%
}{\Phi _0^2}\frac{R_D^2}{R_S},
\end{equation}
where $k_{{\rm B}}$ is Boltzmann's constant, $T$ is the temperature. The
linewidth depends on the differential resistance $R_D=dV/dI$ at the junction
bias point and the static resistance $R_S=V/I$, where $I$ is the bias
current flowing through the junction.

There are two main regimes which characterize the fluxon motion in long
junctions. First, a shuttle-like {\em resonant fluxon motion} gives rise to
zero-field steps (ZFSs) in the dc current-voltage ($I-V$) characteristics of
the junction. In this regime fluxons and antifluxons undergo reflections
from the junction boundaries and the radiation frequency is determined by
the junction length $L$ and the fluxon velocity $v$ as $f_{{\rm ZFS}}=v/2L$
. Second, in the high magnetic field, the so-called {\em flux-flow regime}
occurs and is manifested by a flux-flow step (FFS) on the $I-V$ curve. In
this regime fluxons are created at one boundary of the junction and
annihilate at the other boundary. The radiation frequency $f=v/d_{{\rm fl}}$
is determined by the spacing between moving fluxons $d_{{\rm fl}}$. In
general, for both ZFS and FFS regimes, the radiation linewidth $\delta f$
should be related to thermal fluctuations of the fluxon velocity $v$.
Joergensen et al. \cite{Joergensen:82} obtained a striking general result
that, in spite of a different nature of the phase slippage in small and long
junctions, the linewidth of the resonant single-fluxon radiation in a long
junction is given by the same Eq.~(\ref{JJ-linewidth}), except for a missing
factor of $4$ due to the modified Josephson relation $f=V/2\Phi _0$ at the
ZFS. Experiment \cite{Joergensen:82} showed a reasonable agreement with that
theory, though some excess linewidth broadening has been seen. For the
multi-fluxon state which forms FFS it can be argued that internal degrees of
freedom in the moving fluxon chain may yield a significant contribution in $%
\delta f$. Here, in contrast to the resonant single-fluxon case, local
variations of the fluxon spacing $d_{{\rm fl}}$ change the radiation
frequency. Recent FFS radiation linewidth measurements by Koshelets et al. 
\cite{Koshel-linewid:PRB95} showed the scaling of $\delta f_{{\rm FF}}$ as
predicted by Eq.~(\ref{JJ-linewidth}) but with an effective temperature $T_{%
{\rm eff}}$ being by factor of $8$ larger than the physical temperature of
their experiment. Since there was no theory for the radiation linewidth $%
\delta f$ in the flux-flow regime, the reason for the excess noise was not
resolved.

In this paper we present a theoretical model for the radiation linewidth in
the flux-flow regime for long Josephson junctions. We start from the
perturbed sine-Gordon model with the spatially and temporarily dependent
noise current. Using the collective coordinate approach we calculate the
finite radiation linewidth due to the internal degrees of freedom in the
moving fluxon chain. The obtained analytical result is evaluated for the
relevant parameter range of experimentally studied flux-flow oscillators.

First, we introduce main characteristic quantities for the system. Mean
frequency of Josephson oscillations for a long junction in the flux-flow
regime is 
\begin{equation}
\label{MF}\left\langle f\right\rangle =\frac{v\left\langle H\right\rangle
\Lambda }{\Phi _0}\equiv \beta \left\langle H\right\rangle , 
\end{equation}
where $v$ is the velocity of the fluxon chain, $\left\langle H\right\rangle $
is the average magnetic field, $\Lambda =2\lambda_L+t$ is the effective
magnetic thickness of the junction ($\lambda _L$ is the London penetration
depth, and $t$ is the insulator thickness). The quantities essential for the
linewidth problem are the mean square deviation of the frequency $f$

$$
\sqrt{\left\langle \delta f^2\right\rangle }=\beta \sqrt{\left\langle \left[
H-\left\langle H\right\rangle \right] ^2\right\rangle }=\beta \sqrt{
\left\langle H^2\right\rangle -\left\langle H\right\rangle ^2} 
$$
and reduced r.m.s. linewidth in the flux-flow regime

\begin{equation}
\label{LW}\Delta f_{FF}\equiv \frac{\sqrt{\left\langle \delta
f^2\right\rangle }}{\left\langle f\right\rangle }=\sqrt{\frac{\left\langle
H^2\right\rangle }{\left\langle H\right\rangle ^2}-1}\text{ .} 
\end{equation}
In addition to the thermal fluctuations of the fluxon velocity \cite
{Joergensen:82}, there exists an additional mechanism of the line broadening
in the FFS regime. According to Eq.~(\ref{LW}), this mechanism is directly
related to the irregularities of magnetic field distribution in the
junction, i.e. to the fluxon density fluctuations in the moving fluxon
chain. The most general physical origin for these fluctuations is the
chain's deformation under the influence of thermal noise. Therefore, the
thermal noise causes two different contributions to the total linewidth in
the FFS regime. In the following, the dependence of the fluxon chain
deformations on the velocity and magnetic field and their contribution to
the linewidth are calculated.

The magnetic field distribution in the junction is proportional to the
gradient of the phase difference $\phi (x,t)$ between superconducting
electrodes: $H(x,t)=(\Phi _0/2\pi \Lambda \lambda _J)\phi _x(x,t)$, where $%
\lambda_J$ is the Josephson penetration depth. In the following we note the
normalized magnetic field as $h(x,t)\equiv\phi_x(x,t)$. The phase evolution
is governed by the perturbed sine-Gordon equation (written in the laboratory
coordinates) \cite{McL-Sc:78}

\begin{equation}
\label{SinG}\phi _{xx}-\phi _{tt}-\sin \phi =\gamma +\alpha \phi _t+n(x,t) 
\end{equation}
where the length $x$ and time $t$ are normalized to $\lambda _J$ and to the
inverse plasma frequency $\omega _p^{-1},$ respectively. The first term in
the right hand side of Eq.~(\ref{SinG}) $\gamma $ represents the external
bias current density $J$ normalized to the critical current density $J_c$, $%
\alpha $ is the damping coefficient, and the term $n(x,t)$ represents the
thermal fluctuations \cite{Joergensen:82} with the white-noise
spatio-temporal correlator \cite{B-L:81}

\begin{equation}
\label{Cor}\left\langle n(x_1,t_1)n(x_2,t_2)\right\rangle =\frac{16\alpha
k_BT}{E_o}\delta (x_1-x_2)\delta (t_1-t_2)\equiv \alpha \tau \text{ }
\delta(x_1-x_2)\delta (t_1-t_2). 
\end{equation}
Here the brackets $\left\langle ...\right\rangle $ stand for the statistical
averaging (over the ensemble), and $E_0=8\hbar J_cW\lambda _J/2e$ is the
rest energy of a fluxon, $W$ is the junction's width and $e$ is the electron
charge.

The phase distribution in the flux-flow regime is given by the cnoidal-wave
solution written in the laboratory coordinates as 
\begin{equation}
\label{modul}\phi =\phi _{\text{cn}}(x-vt-\xi ),\text{ where }\phi _{\text{%
cn }}(z)=\pi -2\text{am}(z/k\sqrt{1-v^2}). 
\end{equation}
Here am($z$) is the elliptic amplitude, $v$ is the velocity of the fluxon
chain, and the slowly varying shift $\xi $ describes the chain's
deformation. In the coordinate frame moving at the velocity $v$, $x^{\prime
}=(x-vt)/\sqrt{1-v^2}$ and $t^{\prime }=(t-vx)/\sqrt{1-v^2},$ the
deformation is governed by the linear equation \cite{Supersol}

\begin{equation}
\label{Eq}\xi _{t^{\prime }t^{\prime }}-\xi _{x^{\prime }x^{\prime }}+\alpha
\xi _{t^{\prime }}=\rho ^{-1}n(x^{\prime },t^{\prime }),
\end{equation}
where $\rho =4E(k)/k^2K(k)$ is the density of the fluxon chain, the elliptic
modulus $k$ is related to the chain's period $l^{\prime }=2kK(k)$ in the
moving frame, $K(k)$ and $E(k)$ being the complete elliptic integrals of the
first and the second kind, respectively. The average velocity of the fluxon
chain $v$ under the external bias current $\gamma $ is given by the Marcus
and Imry perturbation approach \cite{Marcus-Imry} as

\begin{equation}
\label{Vel}v=\left[ 1+\left( \frac{4\alpha }{\pi \gamma }\frac{E(k)}k\right)
^2\right] ^{-1/2}.
\end{equation}
This relation determines the form of the current-voltage ($I-V$)
characteristics in the flux-flow regime. In the laboratory coordinate frame $%
k$ in Eq.(\ref{Vel}) is determined by given fluxon spacing $l$ as a root of
the transcendental equation $l/\sqrt{1-v^2}=2kK(k)$ \cite{Marcus-Imry}.

The local magnetic field in the junction $h(x,t)$ is given by the expression

\begin{equation}
\label{H}
\begin{array}{c}
h(x,t)=\phi _{\text{cn}}^{^{\prime }}(x-vt-\xi )(1-\xi _x)\simeq \left[ \phi
_{\text{cn}}^{^{\prime }}(x-vt)-\phi _{\text{cn}}^{^{\prime \prime
}}(x-vt)\xi \right] \left( 1-\xi _x\right) , 
\end{array}
\end{equation}
where, using the smallness of $\xi ,$ we have Taylor expanded the first
multiple in Eq.~(\ref{H}), keeping the first two terms. Next we define the
mean and mean squared magnetic fields as

\begin{equation}
\label{Df}\left\langle h\right\rangle =\phi _{\text{cn}}^{^{\prime }}(x-vt)%
\text{, }\left\langle h^2\right\rangle =\left\langle \left( \phi _{\text{cn}%
}^{^{\prime }}-\phi _{\text{cn}}^{^{\prime \prime }}\xi -\phi _{\text{cn}%
}^{^{\prime }}\xi _x\right) ^2\right\rangle . 
\end{equation}
Substituting (\ref{Df}) into Eq.~(\ref{LW}), we obtain the following
relation between the reduced r.m.s. linewidth and the deformation of the
fluxon chain $\xi $:

\begin{equation}
\label{LW2}\Delta f_{FF}=\sqrt{\left\langle \xi _x{}^2\right\rangle
+\left\langle \left[ \xi \frac{\phi _{\text{cn}}^{^{\prime \prime }}} {\phi
_{\text{cn}}^{^{\prime }}}\right] ^2\right\rangle }. 
\end{equation}
Thus, in order to calculate $\Delta f_{FF}$ we need to find the solution of
Eq.~(\ref{Eq}) which can be done by means of the Fourier transform. First,
we write Eq.~(\ref{Eq}) in the laboratory reference frame:

\begin{equation}
\label{EqL}\xi _{tt}-\xi _{xx}+\frac{\alpha v}{\sqrt{1-v^2}}\xi _x+\frac 
\alpha {\sqrt{1-v^2}}\xi _t=\rho ^{-1}n(x,t).
\end{equation}
Using the Fourier transform $\xi (x,t)=\int_{-\infty }^{+\infty }\xi
(q,\omega )\exp (iqx-i\omega t)$ $dqd\omega $ it is straightforward to find
the solution to Eq.~(\ref{EqL}): 
\begin{equation}
\label{XiF}\xi (q,\omega )=\frac{\rho ^{-1}n(q,\omega )}{q^2-\omega
^2-i\alpha (\omega -vq)(1-v^2)^{-1/2}},
\end{equation}
where $n(q,\omega )$ is the Fourier transform of the thermal noise $%
n(x,t)=\int_{-\infty }^{+\infty }n(q,\omega )\exp (iqx-i\omega t)$ $%
dqd\omega $ which is subject to correlations 
\begin{equation}
\label{Cor1}\left\langle n(q,\omega )n(q^{\prime },\omega ^{\prime
})\right\rangle =\frac{\alpha \tau }{(2\pi )^2}\delta (q+q^{\prime })\delta
(\omega +\omega ^{\prime }).
\end{equation}

Here, we consider only the case of sufficiently dense fluxon chain in the
form 
\begin{equation}
\label{Stiff}\phi _{\text{cn}}\simeq \left\langle h\right\rangle
(x-vt)-\left\langle h\right\rangle ^{-2}(1-v^2)^{-1}\sin (\left\langle
H\right\rangle (x-vt)),
\end{equation}
which assumes

\begin{equation}
\label{vmax}\left\langle h\right\rangle ^2(1-v^2)\gg 1. 
\end{equation}
In this case, the coefficient which enters Eq.~(\ref{LW2}) is ($\phi $$_{%
\text{cn}}^{\prime \prime }/\phi _{\text{cn}}^{\prime })^2\simeq
\left\langle h\right\rangle ^{-2}(1-v^2)^{-2}.$ Note, that according to Eq.~(%
\ref{EqL}) we have $\xi _x^2/\xi ^2\sim \alpha ^2v^2(1-v^2)^{-1}$. It is
easy to see that, if we add to the chain stiffness condition (\ref{vmax})
the additional condition

\begin{equation}
\label{alphamax}\alpha ^2\left\langle h\right\rangle ^2v^2(1-v^2)\ll 1
\end{equation}
the first term under the square root in Eq.~(\ref{LW2}) may be neglected as
compared to the second term. The condition (\ref{alphamax}) holds in most
practically important case of the underdamped junction. Since $v^2(1-v^2)$
is always below its maximum value of $1/4$, the conditions (\ref{vmax}) and (%
\ref{alphamax}) can be combined as

\begin{equation}
\label{hregion}\frac{1}{1-v^2}\ll\left\langle h\right\rangle^2\ll\frac{4}{%
\alpha^2}\:. 
\end{equation}
For the region (\ref{hregion}) we thus obtain

\begin{equation}
\label{LW3}\Delta f_{FF}=\left\langle h\right\rangle ^{-1}(1-v^2)^{-1}\sqrt{
\left\langle \xi ^2(x,t)\right\rangle }\:, 
\end{equation}
where we have inserted the above approximation for ($\phi_{\text{cn}}
^{\prime \prime }/\phi _{\text{cn}}^{\prime })^2$. For the stiff fluxon
chain its elliptic modulus $k$ and mass density $\rho $ are given by
particularly simple expressions $k=2/\left\langle h\right\rangle \sqrt{1-v^2}
$ and $\rho =\left\langle h\right\rangle ^2(1-v^2) $.

Now, what remains is to calculate the r.m.s. value $\sqrt{\left\langle
\xi^2\right\rangle }$. Using Eqs.~(\ref{XiF}) and (\ref{Cor1}) this quantity
can be calculated explicitly. However, we encounter a divergence when
performing integration over small wave numbers $q$. This divergence is
regularized by the fact that, in a Josephson junction of the finite length $%
L $, the smallest wave number is $\pi/L$. Taking this into account we
finally obtain

\begin{equation}
\label{LW4}\Delta f_{FF}=\frac 4{\pi (1-v^2)^{3/2}\left\langle
h\right\rangle ^3}\sqrt{\frac L{\lambda _J}\frac 1\alpha \frac{k_BT}{E_0}}. 
\end{equation}
This expression provides 
the linewidth of a Josephson flux-flow oscillator. It is related to the
intrinsic mechanism of the fluxon chain deformation under the influence of
thermal noise.

The calculation of another noise contribution $\Delta f_T$ due to the
thermal motion of the fluxon chain as a whole (similar to that of Ref.~\cite
{Joergensen:82}) give the result identical to that of Eq.~(\ref{JJ-linewidth}%
). Thus, the total reduced linewidth of the Josephson flux-flow oscillator
is given by the expression

\begin{equation}
\label{LWtotal}\Delta f_\Sigma =\Delta f_T+\Delta f_{FF}.
\end{equation}

Let us discuss the physical meaning of the result (\ref{LWtotal}). First,
the term $k_BT/E_0$ is typically small, of the order of magnitude 10$^{-4}-$%
10 $^{-5}$. Therefore, a comparison of Eq.~(\ref{JJ-linewidth}) with Eq.~(%
\ref{LW4}) shows that, due to the square-root dependence of $\Delta f_{FF}$
on $k_BT/E_0$, the linewidth in the flux-flow regime can be essentially
broader as compared to the resonant flux motion regime. Moreover, the
flux-flow linewidth scales with the junction length as $\sqrt{L}$, which
means additional broadening for longer junctions. Physically, this
dependence is related to a formal divergence of the chain's deformation in
the infinitely long junction.

According to Eq.~(\ref{LW4}), the linewidth is rapidly reduced with the
increase of the magnetic field $h$ which makes the fluxon chain more stiff.
In sufficiently high magnetic fields $h>h^{*}$ the contribution determined
by Eq.~( \ref{LW4}) may become smaller than that due to thermal fluctuations
of the chain velocity given by Eq.~(\ref{JJ-linewidth}). In this case the
crossover to the standard mechanism discussed in \cite{Joergensen:82} takes
place and the total linewidth is determined by Eq.~(\ref{JJ-linewidth}).

The crossover magnetic field $h^{*}$ depends essentially on the
dimensionless ratios $k_BT/\alpha E_0$, $L/\lambda _J$ and on the fluxon
chain velocity $v$. We note that Eqs.(\ref{JJ-linewidth}) and (\ref{LW4})
predict quite different dependencies on $v$. Namely, $\Delta f_T$ decreases
with $v$ and becomes very small in the relativistic regime $v\rightarrow 1$.
In contrast to that, $\Delta f_{FF}$ increases with $v$.

In order to compare $\Delta f_T$ and $\Delta f_{FF}$ it is useful to
estimate them for typical parameters of practical Josephson tunnel
junctions. Figure \ref{linewid:comp} shows the calculated reduced linewidths
given by Eqs.~(\ref{JJ-linewidth}) and (\ref{LW4}) as functions of the
fluxon velocity for two typical sets of experimental parameters \cite
{myASC-92,Koshel-linewid:PRB95}. In order to determine $R_D$ and $R_S$ in
Eq.~(\ref{JJ-linewidth}), the form of the current-voltage ($I-V$)
characteristics for the fluxon spacing $l=2\pi /h$ was calculated using
Marcus and Imry perturbation formula (\ref{Vel}) which with given parameters
was independently checked by numerical simulations for periodic boundary
conditions. In both cases we took the dissipation coefficient $\alpha =0.02$
and the normalized magnetic field $h=4$ (which is approximately twice the
first critical field of the fluxon penetration into the junction) at $T=4.2$%
K. In case of low-$J_C$ junctions \cite{myASC-92} ($J_C=200$A/cm$^2$) we
used $L=200\mu $m, $W=10\mu $m, $\lambda _J=35\mu $m, and in the other
case of high-$J_C$ junctions \cite{Koshel-linewid:PRB95} ($J_C=8000$A/cm$^2$%
) the values $L=450\mu $m, $W=3\mu $m, $\lambda _J=4\mu $m were used. In
Fig.2 one can see that in both cases the effect of the spatially-dependent
fluxon chain noise contribution $\Delta f_{FF}$ given by Eq.~(\ref{LW4}) is
dominating, in particular at high fluxon velocities. At low velocities the
reduced linewidth $\Delta f_T$ (\ref{JJ-linewidth}) is formally diverging
due to finite value of $\delta f_T$ at $f=0$, while the intrinsic flux-flow
linewidth $\Delta f_{FF}$ saturates at the finite level according to Eq.~(%
\ref{LW4}). In both cases the rest fluxon energy $k_BT/E_0$ is of the order
of $\,10^{-5}$J. For the used magnetic field $h=4$ the stiff fluxon chain
assumption (\ref{vmax}) strictly holds only for $v<v_s=0.6$. Since at $v>v_s
$ the fluxon chain can only become more soft, Eq.~(\ref{LW4}) gives the
lowest limit for the expected $\Delta f_{FF}$.

We note, however, that in experiment the flux-flow oscillators are often
operated close to a resonant regime. In such cases the junction frequency
spectrum is modulated by characteristic eigenfrequencies which determine the
oscillation amplitudes (\ref{XiF}). In the resonant states the linewidth can
be expected to be narrowed by cavity resonances (known as Fiske steps) in
the junction. Thus, the non-resonant flux-flow frequency linewidth given by
Eq.~(\ref{LW4}) can be taken as the {\em upper} limit for the real flux-flow
oscillator of the finite length.

Finally, we like to mention that additional mechanism of the flux-flow
oscillator linewidth broadening may arise from technological inhomogeneities
in the junction. Such a contribution is nonuniversal and depends strongly on
the particular type of disorder (local imperfections in the tunnel barrier,
the precision of the photolithography-defined junction width, etc.), and
therefore this mechanism was not discussed in the present paper. This issue
will be the subject of further investigation.

\section*{Acknowledgments}

We would like to thank M.~Cirillo and V.~P.~Koshelets for useful
discussions. One of the authors (B.A.M.) appreciates the hospitality of KFA
J\"ulich. The work was partially supported by INTAS grant no.~94-1783.

\newpage

\begin{figure}
\caption
{Schematic cross-section of a long Josephson junction in
the flux-flow state. The electromagnetic radiation emitted at the
junction boundary depends on the variations of the fluxon spacing due
to temporaly and spatially dependent noise current through the
junction.}
\label{FFS:schematics}
\end{figure}

\begin{figure}
\caption
{Reduced flux-flow oscillations linewidth $\delta f \equiv \Delta f /f$  
given by  Eq.~({\protect\ref{LW4}}) (solid lines) and 
Eq.~({\protect\ref{JJ-linewidth}}) (dashed lines) as a function of the 
fluxon chain velocity $v$. Thick and thin lines correspond to two 
typical parameter sets accounting for Refs.~{\protect\cite{myASC-92}} 
and {\protect\cite{Koshel-linewid:PRB95}}, respectively.}
\label{linewid:comp}
\end{figure}

\end{document}